\documentclass[twocolumn, a4paper, 10pt, showpacs, prx, reprint, aps, superscriptaddress, amsmath, amssymb, amsfonts, floatfix, showkeys]{revtex4-2}

\usepackage{amsmath}
\usepackage{graphicx}
\usepackage{bm}
\usepackage[utf8]{inputenc}
\usepackage{hyperref}
\usepackage{xcolor}
\graphicspath{{figs/}{figsgaoerb/}}

\begin{document}


\title{Magic angle and plasmon mode engineering in twisted trilayer graphene with pressure}

\author{Zewen Wu}
\affiliation{Key Laboratory of Artificial Micro- and Nano-structures of the Ministry of Education and School of Physics and Technology, Wuhan University, Wuhan 430072, China}

\author{Xueheng Kuang}
\affiliation{Key Laboratory of Artificial Micro- and Nano-structures of the Ministry of Education and School of Physics and Technology, Wuhan University, Wuhan 430072, China}

\author{Zhen Zhan}
\email{Corresponding author: zhen.zhan@whu.edu.cn}
\affiliation{Key Laboratory of Artificial Micro- and Nano-structures of the Ministry of Education and School of Physics and Technology, Wuhan University, Wuhan 430072, China}

\author{Shengjun Yuan}
\email{Corresponding author: s.yuan@whu.edu.cn}
\affiliation{Key Laboratory of Artificial Micro- and Nano-structures of the Ministry of Education and School of Physics and Technology, Wuhan University, Wuhan 430072, China}

\date{\today}


\begin{abstract}
Recent experimental and theoretical investigations demonstrate that twisted trilayer graphene (tTLG) is a highly tunable platform to study the correlated insulating states, ferromagnetism, and superconducting properties. Here we explore the possibility of tuning electronic correlations of the tTLG via a vertical pressure. A full tight-binding model is used to accurately describe the pressure-dependent interlayer interactions. Our results show that pressure can push a relatively larger twist angle (for instance, $1.89^{\circ}$) tTLG to reach the flat-band regime. Next, we obtain the relationship between the pressure-induced magic angle value and the critical pressure. These critical pressure values are almost half of that needed in the case of twisted bilayer graphene. Then, plasmonic properties are further investigated in the flat band tTLG with both zero-pressure magic angle and pressure-induced magic angle. Two plasmonic modes are detected in these two kinds of flat band samples. By comparison, one is a high energy damping-free plasmon mode that shows similar behavior, and the other is a low energy plasmon mode (flat-band plasmon) that shows obvious differences. The flat-band plasmon is contributed by both interband and intraband transitions of flat bands, and its divergence is due to the various shape of the flat bands in tTLG with zero-pressure and pressure-induced magic angles. This may provide an efficient way of tuning between regimes with strong and weak electronic interactions in 
one sample and overcoming the technical requirement of precise control of the twist angle in the study of correlated physics. 
\end{abstract}

\maketitle

\section{Introduction}
  Twisted bilayer graphene (tBG) with a “magic angle” ($\backsim 1.1^{\circ}$) has gained extensive attention since the discovery of gate-tunable unconventional superconductivity and strongly correlated insulating phases, which are due to the presence of ultraflat bands near the Fermi level\cite{cao2018unconventional,cao2018correlated}. Recently, experimental and theoretical investigations have been shown that twisted trilayer graphene (tTLG) has a better tunability of its superconducting phases than the twisted bilayer graphene, which makes it a good platform to study the correlated properties\cite{fischer2021unconventional,guerci2021higher,xie2021tstg,zhu2020twisted,lopez2020electrical,cao2021pauli,phong2021band,hao2021electric}. Plenty of degrees of freedom, for instance, the twist angle, stacking configurations, external electric field, and interlayer separation, are available to tune the electronic properties of the tTLG. For example, the interlayer coupling strength can be precisely controlled by the external vertical pressure.

  Previous studies of the pressure effect are mainly focus on bilayer cases. It has been studied experimentally that the hydrostatic pressure can tune the interlayer coupling and hence the band structure of graphene moir\'{e} superlattices\cite{yankowitz2018dynamic,gao2020band}. Interestingly,  for twisted bilayer graphene with a moderate twist angle that shows relatively weak correlation under ambient pressure, an appropriate hydrostatic pressure induces robust insulating phases and superconductivity with higher $T_c$ than that in zero-pressure magic-angle case\cite{feldman2019squeezing,yankowitz2019tuning}. Theoretically, vertical pressure can be used to achieve the ultraflat bands in twisted bilayer graphene with arbitrary twist angle\cite{carr2018pressure,padhi2019pressure,chittari2018pressure,lin2020pressure,ge2021emerging}. Consequently, when studying the strong correlation, we can reduce the impact of structural inhomogeneity by using a moir\'{e} pattern with a short wavelength. However, the vertical pressure effects on the electronic properties of twisted trilayer graphene have not been explored yet. Furthermore, it remains unclear if the flat band tuned by the vertical pressure has similarly peculiar properties as that of the tTLG with zero-pressure magic angle.

Recently, plasmons were detected by utilizing a scattering-type scanning near-field optical microscope (s-SNOM) at tBG with $1.35^\circ$\cite{plasexp2019cao}. Different from the monolayer graphene which displays damped plasmons, tBG with magic angle has collective modes that are damping free. The flat band plasmon modes are ultraflat over the whole wave vector, and with the energy determined by the band width of the flat bands. Such different electronic response to various band widths can be used to identify the magic angle in samples that far beyond the ability of the first-principles methods. Moreover, it has been theoretically predicted that unconventional superconductivity in tBG is mediated by the purely collective electronic modes\cite{sharma2020superconductivity,lewandowski2021pairing}. A deep understanding of the collective excitations in flat band materials, for instance, the tTLG with zero-pressure magic angle and with pressure-induced ``magic angle'', may shed light on the plasmonic superconductivity. Up to now, flat bands are detected in several graphene moir\'{e} superlattices, for example, tBG with magic angle\cite{cao2018unconventional,cao2018correlated}, tBG under moderate pressure\cite{yankowitz2018dynamic,gao2020band}, tTLG with magic angle\cite{fischer2021unconventional,guerci2021higher,xie2021tstg,zhu2020twisted,lopez2020electrical,cao2021pauli,phong2021band,hao2021electric},  trilayer graphene boron-nitride moir\'{e} superlattices\cite{chittari2019gate}, and so on. Natural questions are whether these flat band materials display similar plasmon excitations and the collective modes have similar mechanisms. All in all, a systematic investigation of the collective plasmon modes in graphene moir\'{e} superlattices with flat bands is demanding.

In this paper, we study the pressure effects on the electronic properties of tTLG by means of a full tight-binding (TB) model. We find that an experimental accessible vertical pressure (with the value almost half of that in the tBG case) can push a large twist angle system to reach the flat-band regime\cite{carr2018pressure,feldman2019squeezing,xia2021strong}. Then, the plasmonic properties of the flat band materials are investigated by utilizing the Lindhard function. We observe distinct collective plasmon modes in the tTLG with zero-pressure and pressure-induced magic angles. The outline of the paper is as follows: In Sec. \ref{method}, the TB model and the computational methods are introduced, then followed by the response of the band width and band gap of tTLG-A$\mathrm {\tilde{A}}$A-$1.89^{\circ}$ to the vertical pressure. In sec. \ref{plasm}, we compare the plasmonic properties of the flat band twisted multilayer graphene, in particular, of tTLG-A$\mathrm{\mathrm {\tilde{A}}}$A with zero-pressure magic angle and pressure-induced magic angle. Finally, we give a summary and discussion of our work.

\section{Numerical methods}
\label{method}
\begin{figure}[t]
 	\centering
 	\includegraphics[width=0.48\textwidth]{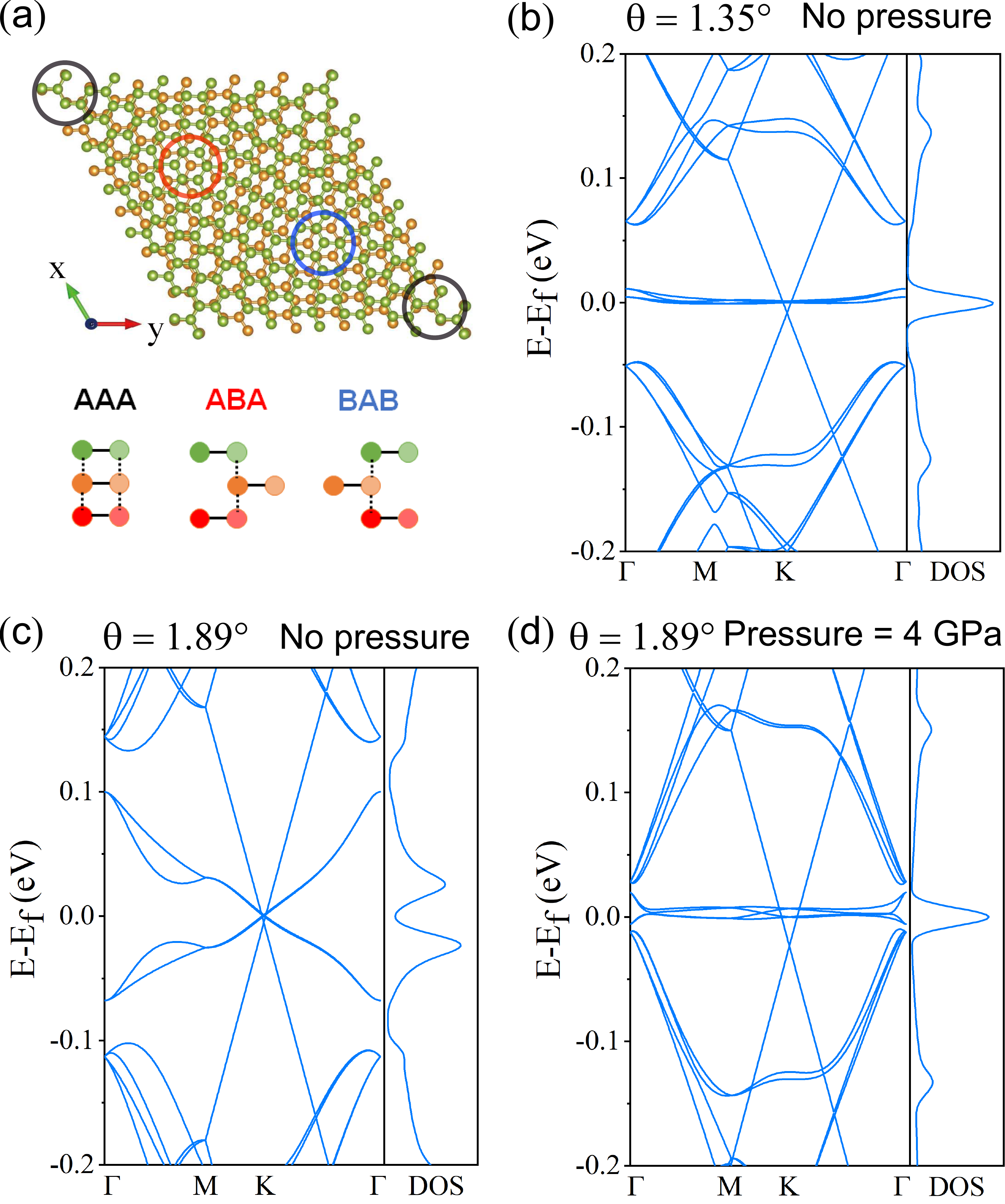}
 	\caption{(a) The upper panel shows the top view of the atomic configuration of tTLG-A$\mathrm {\mathrm {\tilde{A}}}$A-$6.01^{\circ}$. High-symmetry stacking regions of AAA, ABA, BAB are marked by black, red and blue circles, respectively. The lower panel shows a schematic representation of these high-symmetry stacking patterns. The number 6.01 stands for the twist angle $\theta=6.01^\circ$. (b) Band structure and density of states of relaxed tTLG-A$\mathrm {\tilde{A}}$A-$1.35^{\circ}$ with no pressure. (c), (d) Band structure and density of states of relaxed tTLG-A$\mathrm {\tilde{A}}$A-$1.89^{\circ}$ without pressure and with 4 GPa vertical pressure, respectively.}
\label{band_dos}
\end{figure}

The moir\'{e} supercell of twisted trilayer graphene is constructed by identifying a common periodicity between the three layers\cite{shi2020large}. Generally, the electronic properties of the tTLG vary with different stacking configurations\cite{wu2021lattice}. In this paper, we only focus on a mirror-symmetric structure, the so-called tTLG-A$\mathrm{\mathrm {\tilde{A}}}$A, which starts with a AAA stacking ($\theta = 0^\circ$) and with the middle layer twisted an angle $\theta$ with respect to both the top and bottom layers. The rotation origin is chosen at an atom site. As shown in Fig. \ref{band_dos}(a), the supercell is composed of various high-symmetry stacking patterns, that is, the AAA, ABA, and BAB stackings. Moreover, the lattice relaxation (both the out-of-plane and in-plane) is also considered by utilizing the classical simulation package LAMMPS in all calculations\cite{plimpton1995fast}. The intralayer and interlayer interactions in twisted trilayer graphene are simulated with the LCBOP\cite{los2003intrinsic} and Kolmogorov-Crespi potential\cite{kolmogorov2005registry}, respectively.

The electronic properties of the tTLG are obtained by using a full tight-binding model based on $p_{z}$ orbitals. The Hamiltonian of the system has the form\cite{wu2021lattice}:\\
\begin{equation}
H=\displaystyle\sum_{i}\epsilon_i|i\rangle\langle i|+\displaystyle\sum_{\langle i,j\rangle}t_{ij}|i\rangle \langle j|,
\label{ham}
\end{equation}
where $|i\rangle$ is the $p_z$ orbital located at $\mathbf r_i$, $\epsilon_i$ is the on-site energy of orbital $i$, and $\langle i,j\rangle$ is the sum over indices $i$ and $j$ with $i\neq j$.
The hopping integral $t_{ij}$, interaction between sites i and j, is:
\begin{equation}
 t_{ij} = n^2 V_{pp\sigma}(r_{ij}) +(1-n^2)V_{pp\pi}(r_{ij}).
\end{equation}
Here $r_{ij}=|\mathbf r_{ij}|$ is the distance between two sites located at $\mathbf r_i$ and $\mathbf r_j$, n is the direction cosine of $\mathbf r_{ij}$ along the direction $\mathbf e_{z}$ that perpendicular to the graphene layer. The Slater and Koster parameters $V_{pp\sigma}$ and  $V_{pp\pi}$ are expressed as distance-dependent functions\cite{shi2020large}:
\begin{eqnarray}
V_{pp\pi}(r_{ij}) = -\gamma_{0}e^{2.218(b_0-r_{ij})}F_{c}( r_{ij} ),\nonumber \\
V_{pp\sigma}(r_{ij}) = \gamma_{1}e^{2.218(h_0-r_{ij})}F_{c}(r_{ij}),
\label{KS}
\end{eqnarray}
where $b_0=1.42$ \AA\; and $h_0=0.335$ \AA\; represent the nearest carbon-carbon distance and interlayer distance in equilibrium, respectively. The intralayer and interlayer hopping parameters $\gamma_{0}$ = 3.2 eV and $\gamma_{1}$ = 0.48 eV are used in all calculations. $F_{c}(r)=(1 + e^{(r-0.265)/5})^{-1}$ is a smooth function. All the hoppings with $r_{ij}\leq 8.0$ \AA \; are considered in the calculations.

To calculate the electronic properties of the tTLG under a vertical pressure, we extend the hopping parameters in Eq. (\ref{KS}). It has been proven that the vertical compression has a negligible influence on the intralayer interactions, whereas significantly modify the interlayer hoppings\cite{carr2018pressure,lin2020pressure}. Therefore, we only modify the interlayer hopping term $V_{pp\sigma}$ as\cite{lin2020pressure}:
\begin{equation}
V_{pp\sigma}(r_{ij}) = \gamma_{1}e^{2.218(h-r_{ij})}e^{-(h-h_{0})/\lambda^{\prime}}F_{c}(r_{ij}),
\end{equation}
where $\lambda^{\prime} = 0.58$ \AA, h is the out-of-plane projection of $r_{ij}$. The evolution of the lattice constants with the vertical pressure in twisted multilayer graphene has been theoretically investigated, which has the expression as\cite{carr2018pressure,yu2020pressure,gao2020band}:
 \begin{equation}
 \mathrm{Pressure} = A\cdot(e^{B(1 - h/h_0)}-1),
 \label{press}
 \end{equation}
with $A = 5.73$ GPa and $B = 9.54$. Here, $\delta=1-h/h_0$ is the compression.

By direct diagonalization of the Hamiltonian in Eq. (\ref{ham}), we calculate the band structure of tTLG with different twist angles. Moreover, we use the tight-binding propagation method in the frame of a full TB model to calculate the density of states (DOS) as\cite{yuan2010modeling}:
\begin{equation}\label{dos}
D(E)=\frac{1}{2\pi N}\displaystyle\sum_{p=1}^{N}\int_{-\infty}^{\infty}e^{iEt}\langle\varphi_p(0)|e^{-Ht}|\varphi_p(0)\rangle dt,
\end{equation}
where $|\varphi_p(0)\rangle$ is one initial state with the random superposition of basis states at all sites $N$.
To investigate the plasmonic properties of the twisted trilayer graphene, we obtain firstly the dynamical polarization by using the Lindhard function in a full TB model as\cite{kuang2021collective,yuan2011excitation}:
\begin{eqnarray}
 \Pi (\mathbf q,\omega) &&=-\frac{g_{s}}{(2\pi)^{2} }\int_{BZ} d^{2}\mathbf k\sum_{l,l^{\prime}}\frac{n_{F}(E_{\mathbf kl})-n_{F}(E_{\mathbf k^{\prime}l^{\prime}})}{E_{\mathbf kl}-E_{\mathbf k^{\prime}l^{\prime}}+\hbar\omega+i\delta} \nonumber \\
 &&\quad\times|\langle \mathbf k^{\prime}l^{\prime}\mid e^{i\mathbf{q}\cdot\mathbf{r} } \mid \mathbf kl \rangle|^{2}.
\end{eqnarray}
Here, $n_F(H)=\frac{1}{e^{\beta (H-\mu)}+1}$ is the Fermi-Dirac distribution operator, $\beta = \frac{1}{k_BT}$ being $T$ the temperature and $k_B$ the Boltzmann constant, and $\mu$ is the chemical potential; $|\mathbf{k}l \rangle$ and $E_{\mathbf{k}l}$ are the eigenstates and eigenvalues of the TB Hamiltonian in Eq. (\ref{ham}), respectively, with $\mathit{l}$ being the band index, $\mathbf{k^{'}}$=$\mathbf{k}$+$\mathbf{q}$,  $\delta \rightarrow 0^+$, the integral is taken over the whole Brillouin zone (BZ). Then, based on the random phase approximation (RPA), the dielectric function is given by the formula\cite{slotman2018plasmon,jin2015screening,yuan2011excitation}:
\begin{equation}
\varepsilon(\mathbf q,\omega) = 1-V(q)\Pi(\mathbf q,\omega)
\end{equation}
where $V(q)= \frac{2\pi e^{2}}{k\mid q \mid}$ is the Fourier component of the two-dimensional Coulomb interaction, and $\kappa$ is the background dielectric constant. We set $\kappa = 3.03$ to simulate the hexagonal boron nitride  substrate environment in our calculations. Finally, the energy loss function has the form:
\begin{equation}
S(\mathbf q,\omega) = -\mathrm{Im}(1/\varepsilon(\mathbf q,\omega))
\label{loss}
\end{equation}
In principle, undamped plasmons with frequency $\omega_p$ exist if both $\mathrm{Re}\;\varepsilon(\mathbf q,\omega_p)=0$ and the loss function $S(\mathbf q, \omega)$ is peaked around $\omega_p$ with width $\gamma \ll \omega_p$. The loss function can be directly measured by the s-SNOM in the experiment. As a consequence, we will mainly focus on the calculation of the loss function in the paper.

\section{Evolution of bands in twisted trilayer graphene by pressure}

\begin{figure}[t]
 	\centering
 	\includegraphics[width=0.46\textwidth]{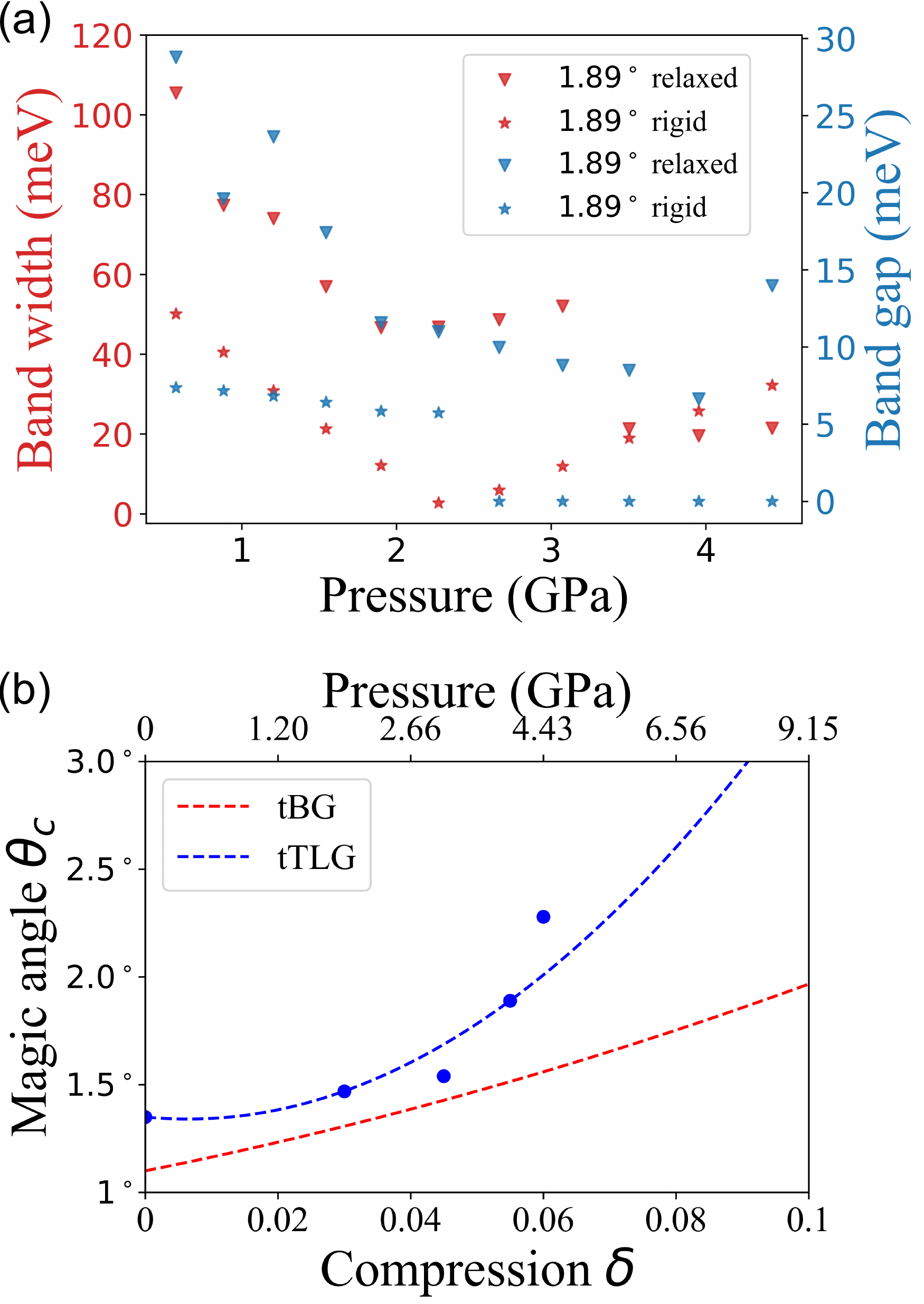}
 	\caption{(a) The band width and band gap of tTLG-A$\mathrm {\tilde{A}}$A-$1.89^{\circ}$ versus the vertical pressure for the rigid and relaxed cases, respectively. The red and blue triangular symbols stand for the band width and band gap in relaxed case, respectively. The red and blue star symbols stand for the band width and band gap in the rigid case, respectively. (b) Pressure-induced magic angle as a function of the critical compression. The blue dashed line is for the tTLG case and the red dashed line is for the tBG case that is extracted from Ref. \cite{carr2018pressure}.}
\label{band_gap_width}
\end{figure}

\begin{figure*}[t]
 	\centering
 	\includegraphics[width=\textwidth]{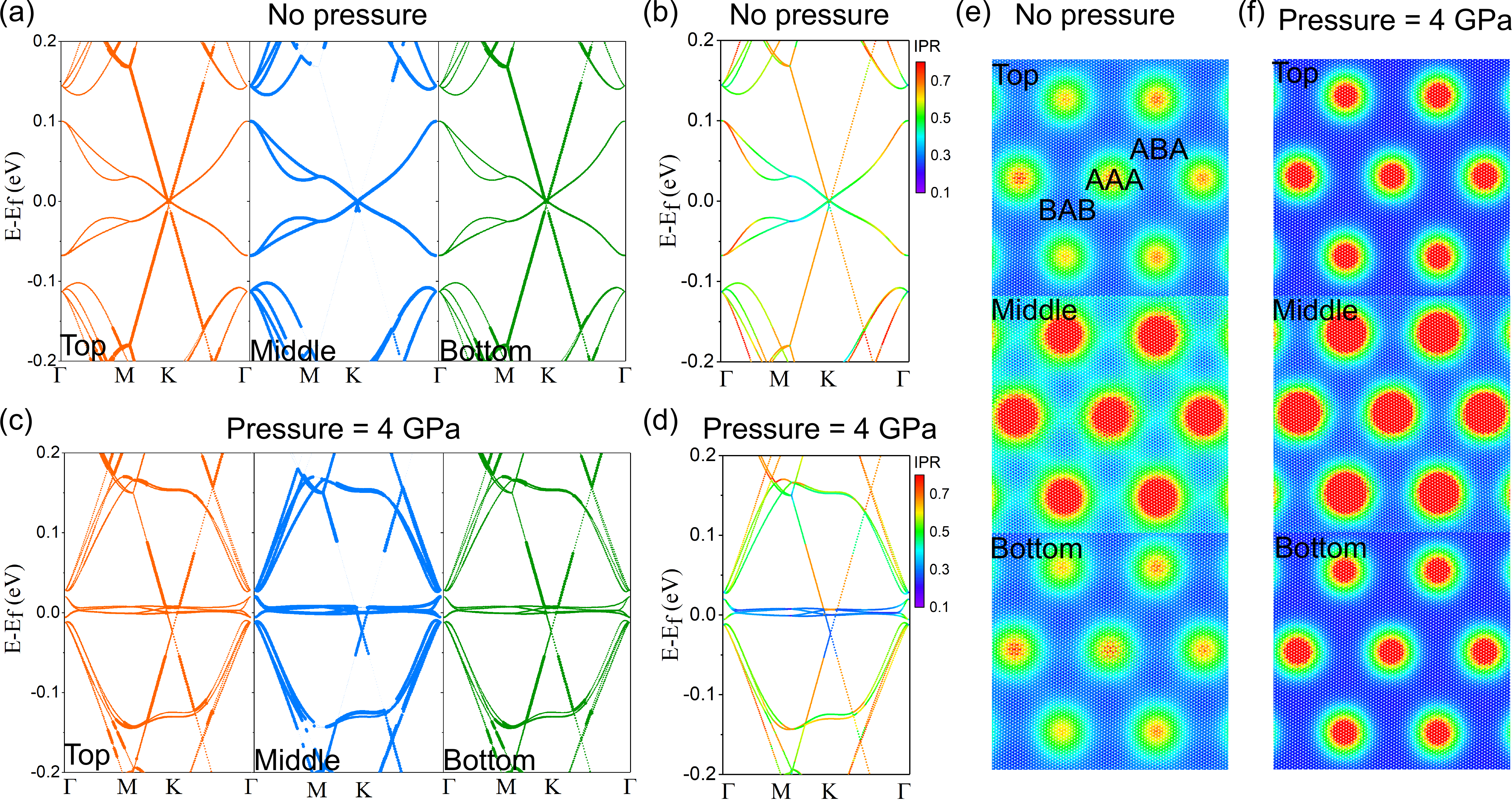}
 	\caption{The electronic properties of relaxed tTLG-A$\mathrm {\tilde{A}}$A-$1.89^{\circ}$ under ambient and high pressures. (a) and (c) The layer-projected weights of band eigenstates of tTLG-A$\mathrm {\tilde{A}}$A-$1.89^{\circ}$ under ambient and 4 GPa vertical pressures, respectively. The thickness of the lines represent the weight of band eigenstates. (b), (d) The band structure and inverse participation ratio (IPR) of the tTLG-A$\mathrm {\tilde{A}}$A-$1.89^{\circ}$ under ambient and 4 GPa external pressure, respectively. (e), (f) Calculated local density of states (LDOS) mappings of Van Hove singularities near the Fermi level of tTLG-A$\mathrm {\tilde{A}}$A-$1.89^{\circ}$ without pressure and with 4 GPa pressure.}
 	\label{OW_IPR}
\end{figure*}

Fig. \ref{band_dos}(b) shows the band structure and density of states of relaxed tTLG-A$\mathrm {\tilde{A}}$A with twist angle $\theta=1.35^{\circ}$ under ambient pressure. The band gap (energy difference between the valence band edge and its higher energy band at the $\Gamma$ point of the Brillouin zone) is about 55.4 meV and the band width (the energy difference between the $\Gamma$ and K points of the valence band edge) is about 10 meV. Similar to the tBG case, the states of four nearly flat bands around the Fermi level show strong localization at the AAA stacking region. One significant difference between the tTLG and tBG is the coexistence of the flat bands with a Dirac cone close to one another only in the mirror-symmetric tTLG-A$\mathrm {\tilde{A}}$A\cite{carr2020ultraheavy}. The relative energy of the Dirac cone with respective to the flat bands is sensitive to the computational parameters of the TB model\cite{wu2021lattice,carr2020ultraheavy}. Here, the Dirac cone is below the flat bands about 18.8 meV. Theoretically, one definition of the ``magic angle'' is the angle where the Fermi velocity at the K and K' points of the BZ vanishes. Another definition is those which lead to the narrowest bands\cite{tarnopolsky2019origin}. In the relaxed tTLG-A$\mathrm {\tilde{A}}$A cases, the narrowest bands appear in sample with $\theta=1.35^{\circ}$--the so-called zero-pressure magic angle. When the twist angle increase to $\theta=1.89^{\circ}$, as illustrated in Fig. \ref{band_dos}(c), the band width is significantly enlarged and the band gap has a value of 48 meV due to the reduced interlayer interactions\cite{yan2012angle}. Moreover, two van Hove singularities flank the Dirac point. Two different sets of linear dispersion bands with different Fermi velocities located at the K point of the BZ. One preserved the monolayer band has Fermi velocity around $9.35*10^{5}$ m/s, and the other has reduced Fermi velocity around $1.53*10^{5}$ m/s due to the interlayer interaction. When applying a vertical pressure with the value of 4 GPa, four nearly flat bands appear near the Fermi level, which can be attributed to pressure-enhanced interlayer correlations. The distortion of the flat bands is different from that of the relaxed tTLG-A$\mathrm {\tilde{A}}$A with zero-pressure magic angle, and the band gap has an obvious decrease. Note that we simulate the pressure by changing the height between two monolayers according to the expression in Eq. (\ref{press}) and the atomic relaxation by allowing the atoms to fully relax in all cases.

\begin{figure*}[t]
 	\centering
 	\includegraphics[width=\textwidth]{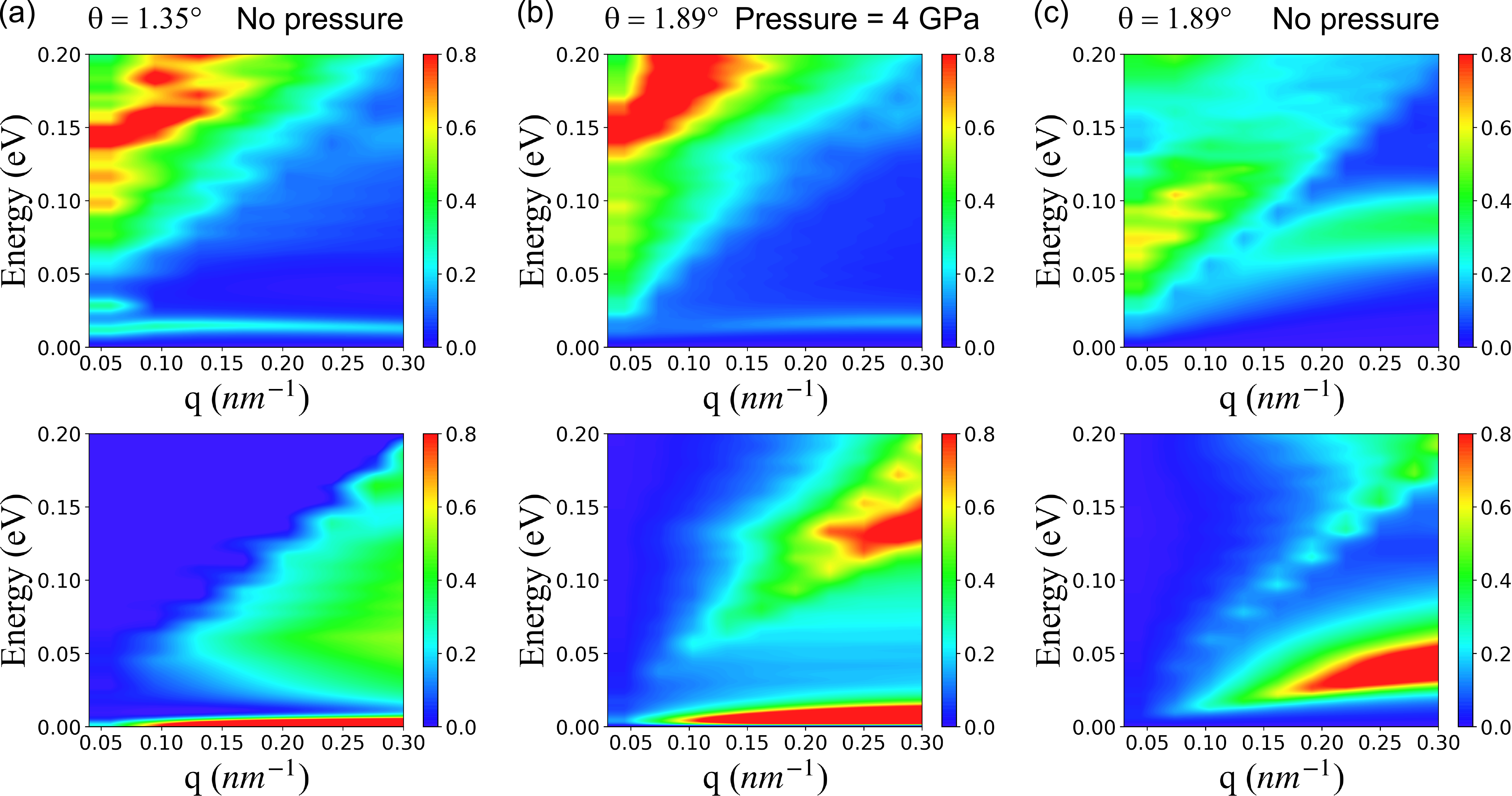}
 	\caption{Plasmonic properties of relaxed tTLG-A$\mathrm {\tilde{A}}$A. The figure is organized in columns. In each column, the upper panel shows the loss function ($-\mathrm{Im}(1/\varepsilon)$) and the lower panel shows the imaginary part of the frequency-dependent dynamic polarization function $-\mathrm{Im}(\Pi(\mathbf q,\omega))$.  (a) Result for tTLG-A$\mathrm {\tilde{A}}$A-$1.35^{\circ}$ under ambient pressure. (b) and (c) Results for tTLG-A$\mathrm {\tilde{A}}$A-$1.89^{\circ}$ under vertical pressure of 4 GPa and under ambient pressure, respectively. The wave vectors is along $\Gamma$ to M in the first Brillouin zone. The temperature is 1 K and chemical potential is $\mu = 0$.}
\label{plasmon}
\end{figure*}

Obviously, similar to the method of precisely controlling the twist angle, pressure is an efficient way of tuning the tTLG-A$\mathrm {\tilde{A}}$A into the magic regime. Next, we investigate how the band structures evolve with the external vertical pressure. As shown in Fig. \ref{band_gap_width}(a), the band gap and band width vary nonmonotonically with the pressure, and such tendency is similar to that of the tBG case\cite{lin2020pressure}. In the rigid sample, the band gap is zero when the pressure is higher than 3 GPa. The band width decreases linearly with the pressure growing up to 2.5 GPa, and then increases linearly after the pressure go beyond the turning point 2.5 GPa, whereas the band gap remains unchanged with the pressure higher than 2.5 GPa. In the relaxed sample, the critical pressure is around 4 GPa, where both the band gap and band width reach their minimum values. That is, for the tTLG-A$\mathrm {\tilde{A}}$A-$1.89^{\circ}$ under 4 GPa vertical pressure, flat-band regime is achieved. Such value of pressure can be achievable experimentally. Recent experimental progress that make use of a hydrostatic pressure allowed to continually tune the interlayer separation in van der Waals heterostructures with pressure up to 2.3 GPa\cite{yankowitz2018dynamic}. Higher pressure would be achievable with diamond anvil cells. By assuming that the interlayer coupling strength has quadratic dependence on compression and neglecting the momentum scattering that the twist angle introduces, we can write the critical value $\theta_c(\delta)$ of the magic angle as a function of compression $\delta$ as\cite{carr2018pressure}:
\begin{equation}
\theta_c(\delta)=\theta_0[(t_2/t_0)\delta^2-(t_1/t_0)\delta+1],
\end{equation}
Here $\theta_0=1.35^\circ$ is the magic angle under ambient pressure, and the numerical parameters are $t_{[0,1,2]}=[1.117, 2.466, 192.496]$. From the results in Fig. \ref{band_gap_width}(b), it is obvious that the pressure needed to induce the flat band in tTLG is lower than  that in the tBG case. Furthermore, the pressure-induced magic angle $\theta_c=3^\circ$ could be achievable when apply a pressure around 10 GPa, where no significant reconstruction appears.

To understand the pressure effect, we compare the electronic properties of tTLG-A$\mathrm {\tilde{A}}$A-$1.89^{\circ}$ with ambient and critical pressures. We calculate the layer-projection weights of band structure of tTLG-A$\mathrm {\tilde{A}}$A-$1.89^{\circ}$ under ambient and moderate pressures in Fig. \ref{OW_IPR}(a) and (c), separately. Firstly, let us focus on the conduction and valence band edges. The middle layer has $50\%$ weight in tTLG-A$\mathrm{\tilde{A}}$A-$1.89^{\circ}$ under zero and critical pressures. The weights are always identical in top and bottom layers, which means that the mirror symmetry is still maintained under pressure. Next, we investigate the localization of the states in tTLG-A$\mathrm {\tilde{A}}$A-$1.89^{\circ}$ with and without critical pressures. The inverse participation ratio (IPR), which is defined as $\sum_{i=1}^N|a_i|^2/(N\sum_{i=1}^N|a_i|^4)$, where $a_i$ is the state at site $i$ and $N$ is the total number of sites, are shown in Fig. \ref{OW_IPR}(b) and (d). After applying a pressure to the tTLG-A$\mathrm {\tilde{A}}$A-$1.89^{\circ}$ sample, the IPR in the conduction and valence band edges change from 0.5 to 0.1, which means the states become more localized. Our LDOS mappings in Fig. \ref{OW_IPR} (e), (f) (the high-symmetry stacking regions are marked by AAA, ABA, BAB in Fig. \ref{OW_IPR}(e)) further clarify the charge concentration process. After applying a vertical pressure, the charges in the  AAA of the top and bottom layer gathering to the AAA center, while in the protected middle layer, charges from the ABA and BAB concentrating to the center of AAA. The charge distribution for the  tTLG-A$\mathrm {\tilde{A}}$A-$1.89^{\circ}$ under critical pressure in Fig. \ref{OW_IPR} (f) is almost the same as magic angle tTLG-A$\mathrm {\tilde{A}}$A-$1.35^{\circ}$ under ambient pressure\cite{wu2021lattice}. All in all, the vertical pressure has similar effect to modify the electronic properties of tTLG as that of tuning the twist angle.

\section{Plasmonic properties of twisted trilayer graphene with magic angles}
\label{plasm}

\begin{figure*}[t]
 	\centering
 	\includegraphics[width=\textwidth]{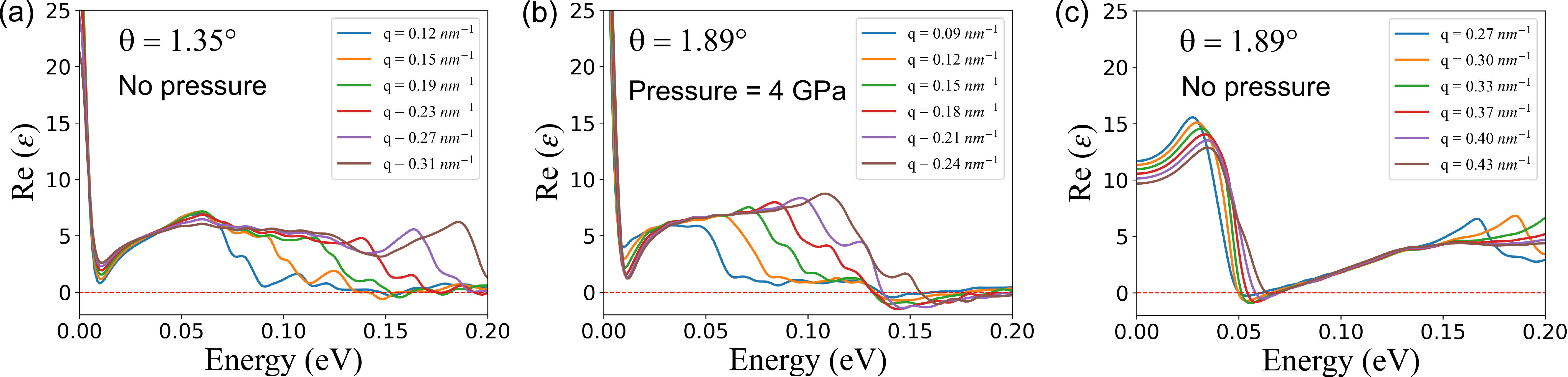}
 	\caption{(a) The real part of the frequency-dependent dielectric function of the relaxed tTLG-A$\mathrm {\tilde{A}}$A-$1.35^{\circ}$ under ambient pressure. (b) and (c) The real part of the frequency-dependent dielectric function of relaxed tTLG-A$\mathrm {\tilde{A}}$A-$1.89^{\circ}$ under 4 GPa and ambient vertical pressures, respectively. The red dashed line indicates the zero of the real part of the dielectric function.}
\label{epsilon}
\end{figure*}

In the previous part, we show that pressure can trigger the appearance of flat bands in tTLG with twist angles larger than the zero-pressure magic angle $\theta=1.35^\circ$. One question arises: will the pressure-induced flat bands show similarly peculiar properties as that of the zero-pressure flat bands? To answer the question, we investigate the plasmon mode in tTLG with magic angles. Generally, when a plasmon mode with frequency $\omega_p$ exists, the electron energy loss spectra possesses a sharp peak at frequency $\omega=\omega_p$. The loss function can be obtained theoretically by using the Eq. (\ref{loss}). For the relaxed tTLG-A$\mathrm {\tilde{A}}$A$-1.35^{\circ}$, as shown in Fig. \ref{plasmon}(a), several collective plasmon modes with energies between 0.01 eV and 0.15 eV appear. Similar to the tBG case\cite{kuang2021collective}, a plasmon mode with energy around 0.15 eV is attributed to the interband transitions from the valence band near the Fermi level to the conduction bands  located at the energy around 0.15 eV. Two collective modes appear in the low energy range. One has an energy of 0.01 eV and stretches to large q. Such plasmon comes from both the interband and intraband transitions of flat bands, and has a weak dependence of the wave vectors, which is due to the interband transition between the flat bands and higher bands\cite{pnasplas2019intrinsically}. The interband transition between the flat bands is suppressed by the interband polarization of flat bands with higher energy bands. This effect is more significant in the second mode with the energy around 0.027 eV, which only appears in small q. The second plasmon is contributed only by the interband transition of flat bands. Such interband polarization is suppressed in the large q range (results not shown here). From the imaginary part of the dynamic polarization functions ($-\mathrm{Im}(\Pi(q,\omega))$) plotted in the bottom panel of Fig. \ref{plasmon}(a), we can see clearly that these plasmon modes are free from Landau damping. The excitons would not exchange energy with  other collective excitations nor have single particle-hole transitions, which means the plasmon mode near 0.15 eV is a long-lived plasmon mode.

For the tTLG-A$\mathrm {\tilde{A}}$A-$1.89^{\circ}$ under critical pressure, a collective plasmon mode locates at energy 0.15 eV, and from the dynamic polarization functions ($-\mathrm{Im}(\Pi(q,\omega))$) in the bottom panel, we can see this kind of plasmon behave the same as the magic angle tTLG-A$\mathrm {\tilde{A}}A-1.35^{\circ}$ under ambient pressure. However, only one plasmon mode appear in the low energy range, which may due to the different shape of the flat bands. Such flat-band plasmon stretches to large wave vector $\bm{q}$, whereas it disperses within particle-hole continuum, as shown in the bottom panel of Fig. \ref{plasmon}(b). This is due to the reduction of the band gap in the pressure-induced magic angle tTLG. For the tTLG-A$\mathrm {\tilde{A}}$A$-1.89^{\circ}$ under ambient pressure, a collective plasmon mode exists around 0.05 eV, and this plasmon mode is a long-lived plasmon mode. The tunable plasmons in Fig. \ref{plasmon} strengthen the finding that a sample with relative larger twist angle can be pushed to the flat-band regime by applying a vertical pressure, which could be justified in experiments by s-SNOM\cite{plasexp2019cao}, electron energy loss spectroscopy\cite{EELS2011electron}.

In principle, there are two different ways to identify the plasmon mode with frequency $\omega_p$. One is the energy where the peak of the loss functions located, which can be seen clearly from the loss functions ($-\mathrm{Im}(1/\varepsilon)$) in the top panel of Fig. \ref{plasmon}; Another way is via identifying the frequencies at which $\mathrm{Re}(\varepsilon(q,\omega))=0$\cite{yuan2011excitation,jin2015screening,slotman2018plasmon}. In Fig. \ref{epsilon}, we plot the frequency-dependent real part of the dielectric functions with varied wave vectors. For the tTLG-A$\mathrm {\tilde{A}}$A-$1.35^{\circ}$ under ambient pressure, the real part of the dielectric functions cross 0 at the energy around 0.15 eV. For the low energy one, there exist dips in the real part of the dielectric functions, and these dips with varied wave vectors approach but never cross the zero dashed line. That means the flat-band plasmon is not a genuine plasmon\cite{nano2016plas}. Such low energy plasmon mode can be tuned to a damping free one by external factors, for instance, an external electric field. Similarly, for the tTLG-A$\mathrm {\tilde{A}}$A$-1.89^\circ$ under critical pressure, the high-energy plasmon mode is free of damping and the low energy flat-band mode damps into particle-hole continuum. The vertical pressure shift the energy of the undamped plasmon in tTLG-A$\mathrm {\tilde{A}}$A$-1.89^\circ$ from 0.05 eV to 0.15 eV. Based on the dielectric properties of the tTLG-A$\mathrm {\tilde{A}}$A with and without pressure, we found that the vertical pressure can push a larger twist angle to reach the flat-band regime, and make their dielectric properties similar to the magic angle tTLG-A$\mathrm {\tilde{A}}$A$-1.35^{\circ}$ under ambient pressure. Furthermore, pressure-induced plasmon mode has a blue shift. That is, a collective of plasmon mode with different energies can be realized continuously by vertical pressure.

\section{CONCLUSION}
We have systematically investigated the evolution of the band widths and band gaps of the tTLG-A$\mathrm {\tilde{A}}$A with an external pressure.  The electronic properties are obtained by employing a full tight-binding model, and the relaxation effects have been taken into account by using the LAMMPS package to fully relax the sample. When applying a vertical pressure with the value around 4 GPa, tTLG-A$\mathrm {\tilde{A}}$A-$1.89^{\circ}$ reach the flat-band regime, both the band gap and band width approach their minimum values. Based on the layer-projected band structure and the LDOS mapping, we found that the appearance of the pressure-induced flat bands is due to the charge concentration in each layer as a result of the enhanced interlayer correlations. The dielectric and plasmonic properties further strengthen our finding that a relatively larger twist angle can be pushed to reach the flat-band regime by vertical pressure. Two plasmonic modes are predicted in tTLG with zero-pressure and pressure-induced magic angles. For the high energy long-lived plasmon, pressure-induced high energy plasmon mode is almost the same as that with zero-pressure magic angle. However, the low energy plasmon mode has obvious divergence, which is probably due to the different shapes of the flat bands in these two kinds of magic angle samples. Recent theory predicts that unconventional superconductivity in TBG is mediated by the purely collective electronic modes\cite{sharma2020superconductivity,lewandowski2021pairing}. This may provide a platform to justify the prediction. Furthermore, we may observe a much higher superconducting $T_c$ in the tTLG with large pressure-induced magic angle\cite{carr2020ultraheavy}. Last but not least, zero-energy high-order van Hove singularity (VHS) has recently emerged as a fascinating playground to study correlated and exotic superconducting phases\cite{yuan2019magic,bi2019designing,guerci2021higher}. Such high-order VHS can be achieved by tuning the band structure with a single parameter in moir\'{e} superlattice, for instance, the twist angle, external pressure, heterostrain and external electric field. It will be worth to explore if a high-order VHS could be induced in tTLG by applying a vertical pressure, which will be our future work.   


\section{ACKNOWLEDGEMENTS}

This work was supported by the National Natural Science Foundation of China (Grants No.11774269 and No.12047543), the National Key R\&D Program of China (Grant No. 2018FYA0305800), and the Natural Science Foundation of Hubei Province, China (2020CFA041). Numerical calculations presented in this paper were performed on the supercomputing system in the Supercomputing Center of Wuhan University.\\

%

\bibliography{references}

\end{document}